\documentclass{aa}

\newcommand\msun{M$_{\odot}$}

\usepackage{graphicx}
\usepackage{amsmath}
\usepackage{natbib}
\usepackage{txfonts}
\usepackage{hyperref}
\usepackage{xcolor}
\begin{document} 
\bibpunct{(}{)}{;}{a}{}{,}

   \title{Tracing the Galactic Disk with Planetary Nebulae using Gaia DR3}
   \subtitle{Distance Catalog, Velocities, Populations, and Radial Metallicity Gradients of Galactic Planetary Nebulae}

   \author{B. Bucciarelli
          \inst{1}
          \and
          L. Stanghellini\inst{2}
          }

   \institute{INAF-OATo,
              Via Osservatorio 20, 10023 Pino Torinese TO, Italy\\
              \email{beatrice.bucciarelli@inaf.it}
         \and
             NSF's NOIRLab, 950 N. Cherry Avenue, Tucson, Arizona 85719, United States\\
             \email{letizia.stanghellini@noirlab.edu}}

   \date{Received; Accepted}

  \abstract
   {}
   {We want to study the population of Galactic planetary nebulae (PNe) and their central stars (CSs) through the analysis of their distances and Galactic distribution. The PN distances are obtained by means of a revised statistical distance scale, based on an astrometrically-defined sample of CSs from Gaia DR3 as calibrators. The new statistical distances, together with the proper motion of the CSs -- also from DR3, and with published PN abundances as well as radial velocities, are used to characterize the PN populations in the Galaxy, and to derive the radial metallicity gradient. }
   {The statistical scale is applied to infer distances of a significant number ($\sim$850) of Galactic PNe, for which we deliver a new catalog of PN distances. By adopting a circular velocity curve of the Galaxy, we also obtain peculiar 3D velocities for a large sample of PNe ($\sim$300). Elemental abundances of the PNe are culled from the literature for an updated catalog, to be used in our analysis and other external applications.}
   {The radial chemical gradient of the Galactic Disk is traced by PNe with available chemical abundances and distances, and kinematic data of the CSs are employed to identify the Halo PN population. 
   We date PN progenitors based both on abundances and on kinematic properties, finding a confirmation of the first method with the second. For all PNe with at least one oxygen determination in the literature, we find a slope of the radial oxygen gradient equal to $\Delta~{\rm log(O/H)}/\Delta~R_{\rm G}=-0.0144\pm0.00385$ [dex kpc$^{-1}$]. Furthermore, we estimate radial oxygen gradients for the PNe with old ($>$~7.5~Gyr) and young ($<$~1~Gyr) progenitors to be respectively $\Delta~{\rm log(O/H)}/\Delta~R_{\rm G}=-0.0121\pm0.00465$ and -0.022$\pm$0.00758 [dex kpc$^{-1}$], thus disclosing a mild steepening of the gradient since Galaxy formation, with a slope change of 0.01 dex. The time evolution is slightly higher ($\sim 0.015$ dex) if we select the best available abundances in the literature. This result is in broad agreement with previous PN results, but now based on DR3 Gaia analysis, and also in agreement with what traced by most other Galactic probes. We also find a moderate oxygen enrichment when comparing the PNe with young and old progenitors.}
   {}


   \maketitle
%

\section{Introduction}

The third data release of the Gaia ESA mission, DR3 \citep{2016A&A...595A...1G,2023A&A...674A...1G} provides unprecedented astrometric -- along with photometric and spectroscopic-- data for hundreds of central stars (CSs) of planetary nebulae (PNe) that, further to witnessing a short-lived phase of their life, turn out to be suitable probes of chemical abundances in the Galaxy \citep[e.g.][]{1978IAUS...76..215P,2010ApJ...724..748H}, as well as in external galaxies \citep[see][and references therein]{2019IAUS..343..174S}. Metal contents of PNe with young and old progenitors can be examined to constrain different mechanisms of gas accretion and star formation history \citep[e.g.][]{2013A&A...554A..47G,2020MNRAS.492..821C}. 

In particular, metallicity gradients along galactic discs are crucial observables in galaxy evolution studies. They are the outcome of complex interacting physical processes involving metal production, consumption, transport, and loss \citep[]{2021A&ARv..29....5M,2023MNRAS.523.2126P}; currently, their radial profiles reconstructed from different metal tracers show some diversities which need to be further investigated \citep[]{2019arXiv190504096S}.
The basic observational step in this direction is to build a metallicity gradient as function of the galactocentric radius $R_{\rm G}$; the profile and slope of the gradient can be inspected for a coeval stellar population, and its time evolution tested with different age populations \citep[e.g.][]{2016A&A...588A..91M,SH18}.
To this end, being able to gauge the precise PN location as function of Galactic radius is clearly of utmost importance. 

In our previous paper \citep[][hereafter Paper I]{PaperII}, we performed a re-calibration of the PN statistical distance scale from a sample of $\approx 100$ CSs having DR2 parallaxes with relative uncertainty better than $20\%$, demonstrating the potential of Gaia astrometry in improving the extant knowledge on PNe distances. Following on this path, we present a revision of the PN distance scale using the best parallaxes available from Gaia DR3 \citep{Lin21}. 
In addition, the highly accurate Gaia DR3 proper motions are precious parameters in the context of this study: in fact, space velocities can be inferred by the CS proper motions using their distances. Since the velocity dispersion of a disk population tends to increase with stellar age \citep[e.g.][]{1925ApJ....61..363S,1946ApJ...104...12S,1977A&A....60..263W,2019MNRAS.489..176M},
the kinematic properties of PNe CSs bear independent information about the progenitor's age, which can complement and support the elemental abundance dating method.
Therefore, in this work we have exploited the full astrometric data provided by Gaia DR3 for a bona-fide set of Galactic PNe, in combination with their most updated chemical abundances and radial velocities, to investigate the gas-phase metallicity gradient of the Milky Way and its evolution in time.

The structure of the paper is as follows. In $\S$ 2 we characterize the Galactic PN sample and give an assessment of the new distance scale calibration, based on their DR3 parallax and other nebular parameters; $\S$ 3 describes the data and methods for the derivation of radial metallicity gradients; finally, $\S$ 4 is devoted to the discussion of the results and the conclusions.

\section{The Revised Distance Scale}
We explore the statistical distance scale for Galactic Planetary Nebulae (PNe) hinging on the commensurate relation between their surface brightness and physical radius advocated by \cite{1956AZh....33..222S}, which has been widely used and refined since then; see \citet{Smith2015} for a thorough review. The similarity between the expansion radius of a fully ionized nebula and the decreasing density of its (nearly) constant ionized mass provides an estimation of the PN distance, once its angular radius and observed flux have been measured. 

Despite the wide range of PNe mass progenitors, the so-called Shklovsky mass, i.e., the one responsible for the observed flux, has a fairly constant value of $\approx 0.1-0.2$ $M_{\odot}$, making the underlying assumption that all nebulae have the same ionized mass a viable one. Nonetheless, the Shklovsky method tends to overestimate the distances of less evolved PNe that are in a "ionization bound" state, i.e., whose ionization front is still within the nebular material, far from the edge of the gaseous cloud \citep{1996ApJ...459..606S}. 

The wealth and unprecedented accuracy of Gaia DR3 parallaxes allow us to calibrate the {\it physical radius-surface brightness} distance scale on well-constrained empirical evidence, while allowing us to put to test some of the crucial aspects of the underlying physical models.
In logarithmic units, the scale is represented by the equation:
\begin{equation}
{\rm log}R= a \times {\rm log}S_{\rm H\beta}+b,
\label{scale}
\end{equation}

where the surface brightness $S_{\rm H\beta}$ is the distance-independent parameter, given in terms of the extinction-corrected absolute $H\beta$ flux from the nebula $I_{\rm H\beta}$ (in erg cm$^{-2}$ sec$^{-1}$) and its angular radius $\theta$ (in arcsec) as $S_{\rm H\beta}=I_{\rm H\beta}/\pi\theta^2$, whereas the physical radius of the PN, $R=\theta"/(206265~p")$ (in pc), with $p$ the PN's parallax, is the distance-dependent variable.
\subsection{Identifying Central Stars of Planetary Nebulae in Gaia DR3}
Starting from a list of 2556 {\it true} PNe extracted from the HASH database \citep{HASH}\footnote{The list includes all objects from the HASH database that were labeled as 'TRUE PN' at the time of this work}, we searched for Gaia DR3 sources with the HASH coordinates of the nebular centers, inside a radius of 3 arcsec and up to 5 arcsec if the smaller radius returned zero matches. We resolved multiple matches by applying a ranking system based on the astrometric quality indicators provided with the Gaia solution. In such a way, we identified 1674 CS candidates lying near the geometric center of the nebula with a DR3 parallax. 

Furthermore, to check the reliability of our purely astrometric criterion, we made comparisons with the CS catalogs recently published by \citet{GS21} and \citet{CW21} whose matching algorithms used the Gaia BP-RP colour to discriminate between hot and cool stars candidates and weigh them accordingly. Since our main aim in this context was sample purity as opposed to completeness, we retained only those objects for which our matched Gaia source was in agreement with both \citet{GS21} and \citet{CW21} catalogs, obtaining a sample of 942 CSs, whose corresponding PNe constitute our {\it Calibrator Sample}.

\subsection{Selection of Distance Scale Calibrators}

To find the parameters {\it a} and {\it b} of Eq.\ref{scale} we matched the Calibrator Sample with the physical parameters that are needed to build the scale. Table~\ref{table:calibrators} contains the Calibrator Sample PNe whose angular radius, $H\beta$ flux, and extinction correction are available in the literature. For each of those targets we list the PN~G name, the Gaia DR3 ID name -- which is different from the DR2 ID, the common name used for the PN, the corrected (see below) DR3 parallax $\varpi_{\rm c}$, the angular radius $\theta$, the logarithmic $H\beta$ flux, the logarithmic extinction constant, and the ionized mass ${\rm log}M_{\rm i}$, the latter calculated following the prescription described in Paper I. For the cases where uncertainties on the physical parameters are unavailable 
we assume a typical error for each parameter. In particular, we adopt 0.2 as a relative angular radius uncertainty, and 0.1 as the uncertainty for the logarithmic extinction. 
\begin{table*}
\caption{PN Calibrators Parameters}          
\label{table:calibrators}      
\begin{tabular}{lllrrrrr}
\hline
    PN~G &              Gaia ID (DR3) &   alias& $\varpi_{\rm c}$&     ${\theta}$&  $F_{\rm H\beta}$&       $c$&   ${\rm log}M_{\rm i}$ \\
&   &   & [$\arcsec$]&   [$\arcsec$]& [erg cm$^{-2}$ s$^{-1}$]& & \\
\hline
000.1+17.2 & 4130784921205604736 &        PC12 & 0.108$\pm$0.081 &   1.1245 & -11.91$\pm$0.00 &  0.700$\pm$0.10 & -1.198 \\
000.1-05.6 & 4049045783774253696 &       H2-40 & 0.361$\pm$0.326 &   8.7950 & -13.20$\pm$0.40 &  0.731$\pm$0.15 & -1.761 \\
000.3+12.2 & 4126115570219432448 &      IC4634 & 0.414$\pm$0.052 &   5.8150 & -10.88 $\pm$0.01 &  0.550$\pm$0.06 & -1.175 \\
\hline
\end{tabular}

\tablefoot{See main text for a detailed column description.}
\tablefoottext{The full table is available electronically at the CDS.}

\end{table*}

All parameters, excluding those derived from Gaia DR3, have been selected from the literature \citep[references in][]{SH18}. Parallaxes in Table~\ref{table:calibrators} are those used to fit the scale, and have been corrected by applying to DR3 parallaxes a global offset of 0.017 mas plus an additional bias term which depends on the star's $G$ magnitude, its colour $G_{BP}-G_{RP}$ --- or {\it pseudo-colour} in case of a 6-parameter (6p) astrometric solution --- and ecliptic latitude, detailed in \citet{2021A&A...649A...4L}. Moreover, parallax errors have been multiplied by 1.05/1.22 for a 5p/6p parameter solution to compensate an underestimation (bias) of the error as described in \citet{Fab21}. In the following, the symbols $\varpi_{\rm c}$, $\sigma_{\varpi_{\rm c}}$ will indicate that the corresponding DR3 parallax $\varpi$ and standard error $\sigma_\varpi$ have been corrected for the above mentioned systematic effects.

When we match our Calibrator Sample with the PNe whose radius, H$\beta$ flux, and extinction are also available, we obtain 401 calibrators. If we chose $\sigma_{\rm \varpi_c}/\varpi_{\rm c}<20\%$ (10$\%$) we still have 137 (74) calibrators, doubling the corresponding samples in the DR2 calibration (Paper I).

\subsection {Building the Planetary Nebula Distance Scale}

The semi-empirical relation expressed by Eq.~\ref{scale} sets the stage for the inference problem relating the sought-for posterior density distribution of the unknown parameters {\it a} and {\it b} to the likelihood of the observed physical quantities characterizing our PN calibrators. We give here a short exposition of the method, referring the reader to the Appendix of Paper I for a complete derivation. 

To formulate our likelihood function, we first introduce the measured radius $\theta \sim N(\phi,\sigma_\theta)$, flux $I \sim N (J,\sigma_I)$, and parallax $\varpi \sim N(p,\sigma_\varpi)$ as Gaussian-distributed uncorrelated variables, then marginalize the likelihood of the i-th set of measurements with respect to the parameters radius ($\phi$), flux ($J$), and parallax ($p$) and treat the parallax as dependent variable through Eq.~\ref{scale}, obtaining the expression of the likelihood for the i-th calibrator as:
\begin{equation}
\begin{split}
L(\varpi_i, \theta_i, I_i|a,b)=\int_{0}^{\phi_{lim}}\int_{0}^{J_{lim}}N(\frac{\pi^a\phi_i^{2a+1}}{206265 J_i^a 10^b},\sigma_{\varpi_i})\times \\
N(\phi_i,\sigma_{\theta_i})N(J_i,\sigma_{I_i})d\phi dJ,
\end{split}
\label{likelihood}
\end{equation}
where we have used uniform priors for $\phi$ and $J$ inside some reasonable intervals.
Then, after forming the {\it log-likelihood} of the complete set of calibrators, by virtue of the Bayes theorem we derive the bi-variate posterior probability density ($p_{\rm pdf}$) of {\it a} and {\it b}, assigning them uniform priors, as:

\begin{equation}
\begin{split}
    p_{\rm pdf}(a,b|\varpi,\theta,I) \propto \sum_{i=1}^{N} \log( \int_{0}^{\phi_{lim}}\int_{0}^{J_{lim}}\left(N(\frac{\pi^a\phi_i^{2a+1}}{206265 J_i^a 10^b},\sigma_{\varpi_i}\right)\times \\
    N(\phi_i,\sigma_{\theta_i})N(J_i,\sigma_{I_i})d\phi dJ).
\end{split}
\label{priors}
\end{equation}
We approach this problem by numerical integration over the $\phi$ and $J$ parameters' range, and by sampling the posterior distribution of {\it a} and {\it b} inside a bi-dimensional grid with step sizes of 0.0005 and 0.005 for the {\it slope} and {\it intercept} parameters, respectively.
The estimated $\hat{a}$ and $\hat{b}$ correspond to the mode of the discrete posterior probability density function. The $68\%$ confidence interval for the two parameters are calculated separately, by projecting onto the {\it a} and {\it b} axes the iso-probability contour of the posterior distribution for which $|\text{p}(a,b)_{\text{max}}-\text{p}(a,b)| \leq 0.5$ (remembering that the log-likelihood ratio statistics is asymptotically distributed as $\chi^2/2$). And finally, we estimate the correlation term from the complete posterior sample, as $\rho_{\hat{a}\hat{b}}=1/(\sigma_{\hat{a}}\sigma_{\hat{b}})\sum_{i=1}^{n_a}\sum_{j=1}^{n_b}[p(a_i,b_j)((a_i-\hat{a})(b_j-\hat{b})]$, obtaning in all cases a high positive correlation between the estimated slope and intercept, as illustrated in Fig.~\ref{fig:postprob} showing density levels of the bi-variate posterior distribution $p_{\rm pdf}(a,b)$ for the $\sigma_{\varpi_{\rm c}}/\varpi_{\rm c}<20\%$ sample.

\begin{figure}
   \centering
   \includegraphics[width=8cm]{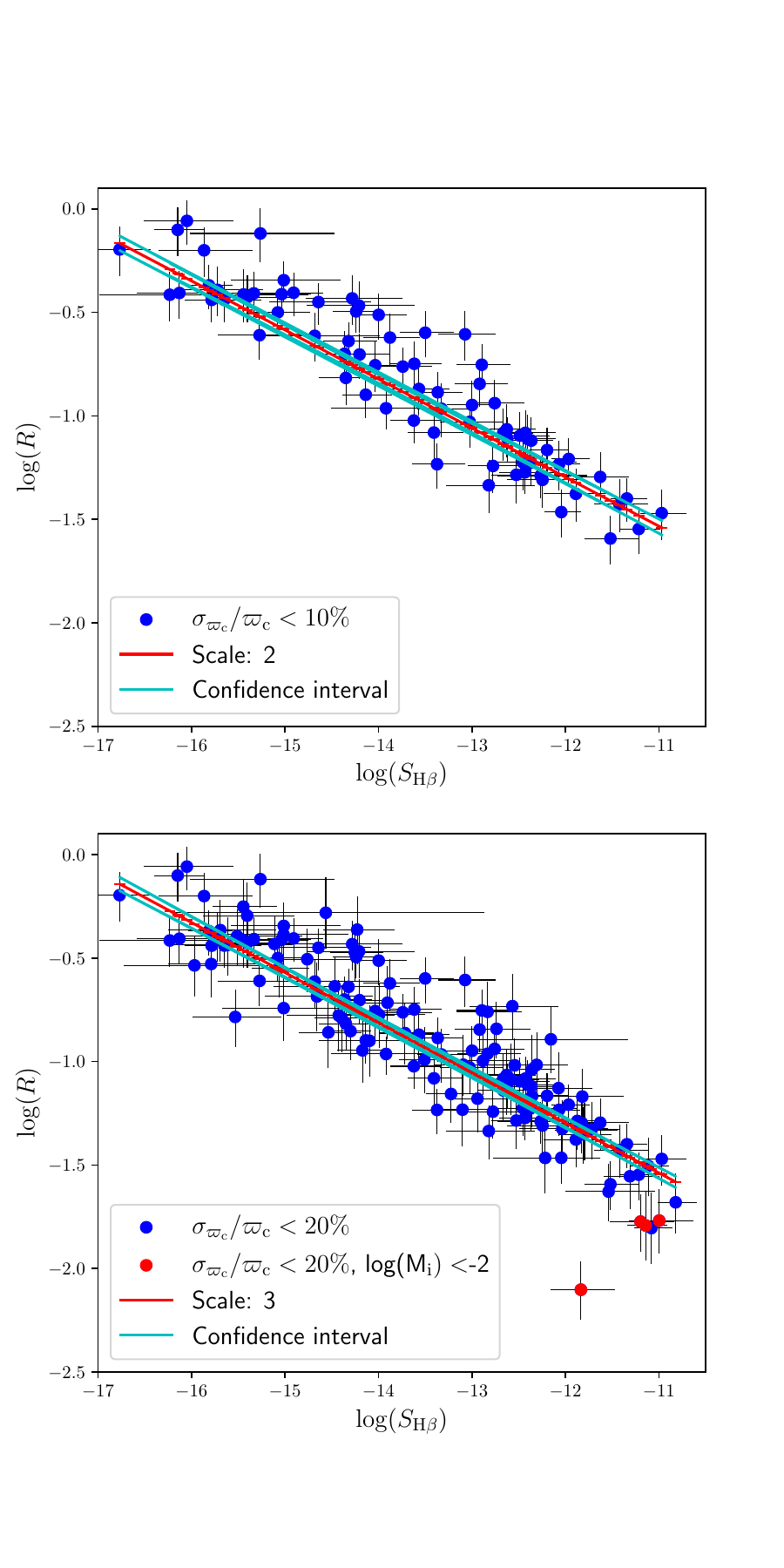} 
   \caption{The log($R$) [pc] vs. log($S_{\rm H\beta}$) [erg cm$^{-2}$ s$^{-1}$] plot for calibrators with $\sigma_{\rm \varpi_c}/\varpi_{\rm c}<0.1$ (top panel) and $\sigma_{\rm \varpi_c}/\varpi_{\rm c}<0.2$ (bottom panel). 
   In both plots, the blue points are the calibrators, with their logarithmic (asymmetric) error bars reflecting the 1-$\sigma$ uncertainties of the observed parameters; In the bottom plot the red points are those calibrators with log$M_{\rm i}<-2$; the red line are the two distance scales from the calibrators, and in cyan we plot the 2-$\sigma$ confidence band of the scales, representing the uncertainty of the logarithmic radius $R$ induced by the correlated errors on $\hat{a}$ and $\hat{b}$}
    \label{fig:scaleplots}
    \end{figure}

We inferred the distance scale parameters using different sets of calibrators, defined by the adopted threshold on the relative DR3 parallax error $\sigma_{\varpi_{\rm c}}/\varpi_{\rm c}$, obtaining consistent estimations of the scale slope {\it a} and intercept {\it b} as reported in Table~\ref{table:smith}. 
Figure~\ref{fig:scaleplots} shows the ${\rm log}R-{\rm log}S_{\rm H\beta}$ plane with the subsets of calibrators for which $\sigma_\varpi/\varpi \leq 10\%$ (top panel) and $\sigma_{\varpi_{\rm c}}/\varpi_{\rm c} \leq 20\%$ (bottom panel) along with their error bars, and the corresponding estimated linear relation (see caption).

By inspecting these plots, both data sets show a somewhat large dispersion around the scale.   
Moreover, from a comparison of the $20\%$ sample of Figure~\ref{fig:scaleplots} with that of  Fig.~2 in Paper I (where the calibration used DR2 parallaxes) we note that the numerous "stragglers" populating the lower right end of the plane, which corresponded to PN with very low ionized mass (log$M_{\rm i}<-2$), are not so prominent in the new plots. 

It turned out that, among the complete set of potential calibrators, 37 objects had log$M_{\rm i}<-2$ but none of them had accurate enough parallaxes at the $10\%$ level and a few -- indicated in red in the lower panel of Figure~\ref{fig:scaleplots}-- met the $20\%$ requirement. If we put these 37 objects on the ${\rm log}R-{\rm log}S_{\rm H\beta}$ plane, they are scattered mainly in the lower part of the graph and their distance do not conform with the statistical scale, corroborating the findings of our previous paper based on DR2 data.  

\begin{figure}
   \centering
   \includegraphics[width=8cm]{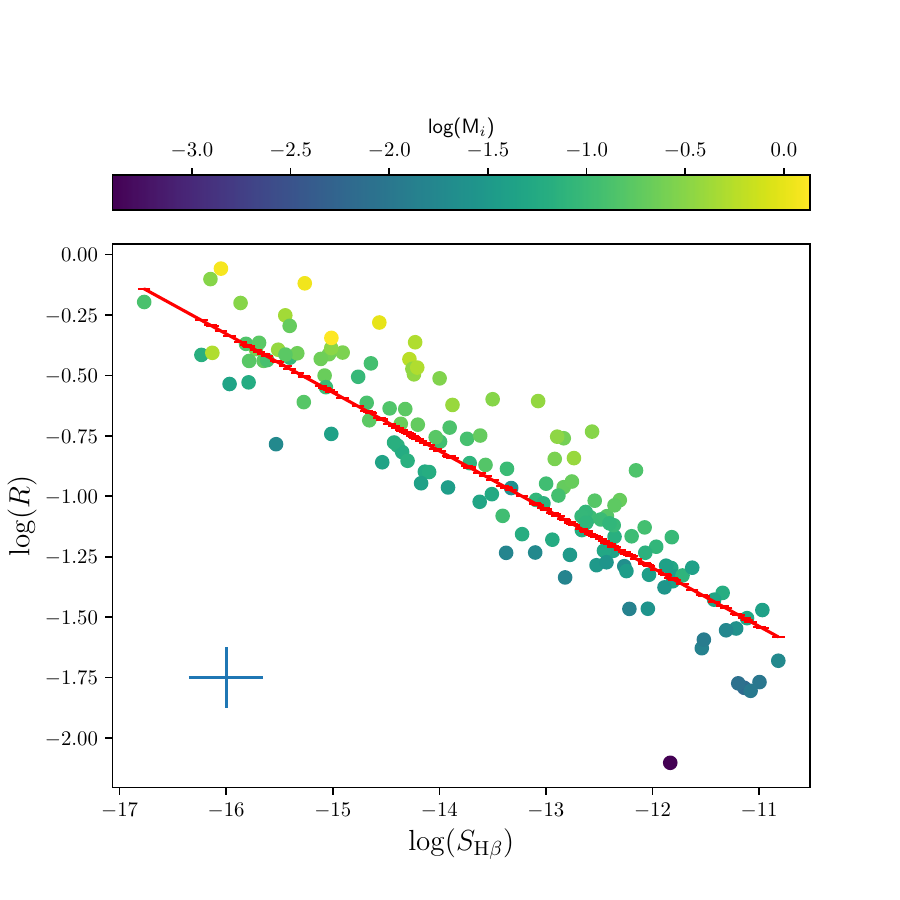} 
   \caption{The log($R$) [pc] vs. log($S_{\rm H\beta}$) [erg cm$^{-2}$ s$^{-1}$] plot for the PN sample with $\sigma_{\rm \varpi_c}/\varpi_{\rm c}<0.2$, colour-coded according to their ionized mass. The error bars have been omitted for clarity; the cross at the bottom-left corner is representative of the typical error in the respective axis}
  \label{fig:logMi}
    \end{figure}

\begin{figure}
   \centering
   \includegraphics[width=8cm]{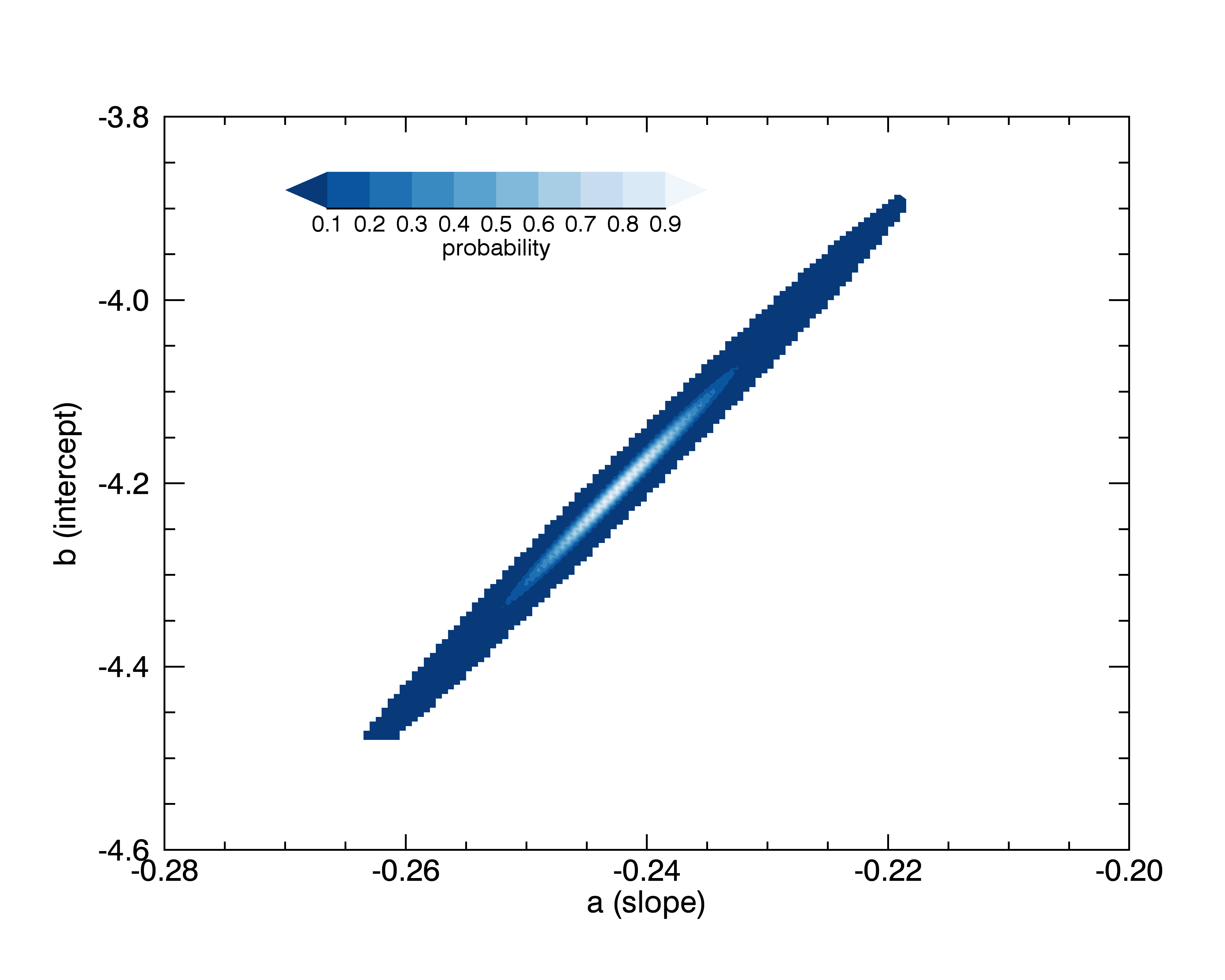} 
   \caption{Normalized 2D posterior probability of the distance scale parameters obtained from the PN sample with $\sigma_{\varpi_{\rm c}}/\varpi_{\rm c}< 20\%$}
  \label{fig:postprob}
    \end{figure}

To further investigate the ionized mass effect, we inspect the PNe with $\sigma_{\rm \varpi_c}/\varpi_{\rm c}<20\%$
by plotting them using a colour-density scale that increases with log$M_{\rm i}$. The result, shown in Figure~\ref{fig:logMi},
is quite striking and can be fully appreciated thanks to the quality of the calibrators: the dispersion around the linear scale is clearly linked to the (estimated) log$M_{\rm i}$ of the PN shell, manifesting the deviation of its actual evolutionary state. In fact, in the transition from ionization-bound, or optically thick, to density-bound, or optically thin, the ionized mass of PNe increases, as discussed in \cite{1980A&A....89..336P}; therefore, the model assumption of fully ionized, constant mass on which Eq.~\ref{scale} stands, does not strictly hold. 

Remarkably, when we tried to tighten even more the constraint on the goodness of the parallax relative error we obtained a similar dispersion of the calibrators as in Fig.~\ref{fig:logMi}. Finally, by subtracting out the effect due to the nominal uncertainties of the observed angular radius and flux, we tentatively estimated an average intrinsic (cosmic) scatter of $\sim 0.1$ dex around the distance scale and conclude that this is due to an actual observed nebular evolution which limits the potential accuracy of this statistical distance scale, unless the ionized mass for the PN is known a priori.

The results of the calibration are reported in Table~\ref{table:smith}, where for each of the three relative uncertainty thresholds on $\sigma_{\varpi_{\rm c}}/\varpi_{\rm c}$ we give the number of calibrators ${\rm N_{cal}}$, the estimated slope and intercept along with their correlation $\rho_{\rm ab}$; we also list the averaged parameter $<K>=<{\tt statistical\_scale\_distance}\times \varpi_{\rm c}>$, ideally equal to 1, and its dispersion $<\sigma_{\rm K}>$, which is an indicator of the goodness of the scale, see \citep{Smith2015}.

\begin{table*}
\caption{Summary of the Distance Scale Analysis}        
\label{table:smith}  
\begin{tabular}{llrrrrrrr}
\hline
Scale & $\sigma_{\rm \varpi_c}/\varpi_{\rm c}$&   $N_{\rm cal}$&  Slope&              Intercept&  $\rho_{ab}$ & $N_{\rm comp}$ &   $<K>$&      $<\sigma_{\rm K}>$\\

\hline
1&$<$0.05&   26&   -0.232$\pm$0.010&   -4.075$\pm$0.132&  $0.99$    & 26 (1) &        0.894&                   0.102\\
&&&&&  &                                                 355 (2)&        0.913&                   0.544\\
&&&&&&&\\
2& $<$0.10&   74&   -0.237$\pm$0.005&   -4.14$\pm$0.070&  $0.99$  &     74 (1)&                                      0.914&      0.113\\
&&&&& &355 (2)&  0.917& 0.543\\
&&&&&&&&\\
3& $<$0.20&   137& -0.242$\pm$0.004&   -4.20$\pm$0.057&  $0.99$  &     133 (1)&                                        0.964&      0.154 \\
&&&&&  &355 (2)&  0.931& 0.553\\
\hline
\end{tabular}
\tablefoot{
$N_{\rm cal}$=number of calibrators; $N_{\rm comp}$ = number of PNe used to compute $<K>$ and $<\sigma_{\rm K}>$:\\
in group (1) we used the Calibrator Sample with the additional requirement of log$M_{\rm i}$>-2, and 
in group (2) we included all PNe with a distance and a parallax determination, and again log$M_{\rm i}$>-2.}

\end{table*}
In this paper we will use the third scale listed in Table~\ref{table:smith}, the one calibrated with central stars whose DR3 parallaxes were measured with better uncertainty than 20$\%$, which has a $<K>\sim0.964$. Through the analysis presented in this paper we have confirmed that using another of the three scales would not have an impact of the science results.
Our adopted scale is:
\begin{equation}
  {\rm log}R_{\rm PN}=(-0.242\pm 0.0042) \times {\rm log}S_{\rm H\beta} - (4.2\pm0.057).
  \label{adopted}
\end{equation}

In Table~\ref{table:catalog} we give the new catalog of distances to 843 Galactic PNe, based on Eq.~\ref{adopted}. For each PN in the HASH catalog whose $H\beta$ intensity and apparent radius is available we calculate its heliocentric distance, $D$, based on our scale.

\begin{table*}
\caption{Catalog of Statistical Distances and Peculiar Velocities of Galactic PNe}          

\label{table:catalog}     
\begin{tabular}{lrrrrrl}
\hline
PN~G&  ${\theta}$&  $F_{\rm H\beta}$&       $c$&   $D$&    $V_{\rm pec}$&   Pop.\\
&  [$\arcsec$]& [erg cm$^{-2}$ s$^{-1}$]& & [kpc]& [km s$^{-1}$] \\
\hline
000.1+17.2 &  1.12 & $-11.91\pm0.00$ &  $0.70\pm0.10$ &  $8.341^{+0.689}_{-0.636}$ & $160.21\pm28.83$ &  Bulge \\
000.2-01.9 &  4.47 & $-12.56\pm0.01$ &  $2.04\pm0.10$ &  $2.793^{+0.229}_{-0.211}$ &  $50.95\pm12.12$ &   Disk \\
000.3+12.2 &  5.82 & $-10.88\pm0.01$ &  $0.55\pm0.06$ &  $2.188^{+0.137}_{-0.048}$ &  $48.23\pm9.72$ &   Disk \\
\hline
\end{tabular}

\tablefoot{Distances are obtained from Scale number 3 of Table ~\ref{table:smith}; see main text for a detailed column description.}
\tablefoottext{The full table is available electronically at the CDS.}

\end{table*}

In the catalog we list the PN~G name, the angular radius $\theta$, the 
logarithmic $H\beta$ flux and its uncertainty, the logarithmic extinction 
constant and its uncertainty, the heliocentric distances from our scale of 
Eq.~\ref{adopted}, $D$ [kpc], and their left/right formal uncertainties. 
The latter have been computed by estimating $\sigma_{\log D}$ via first-order error propagation of the linear relations among the variables 
log$R$, log$S_{\rm H\beta}$ $\theta$ and $D$, accounting for uncertainties in both observed and fitted parameters; then, asymmetric uncertainties were defined by putting $\sigma_{\rm D^-}=10^{(\log D-\sigma_{\log D})}$ and $\sigma_{\rm D^+}=10^{(\log D+\sigma_{\log D})}$. 
Finally, we include the PN peculiar velocity $V_{\rm pec}$ and its uncertainty calculated from Eq.~\ref{eq:peculiar}, plus a Disk/Bulge/Halo population classification as discussed in the following section.
\section{Planetary Nebulae as tracers of the Galactic Disk}

An important scientific motivation for the Galactic PN distance scale is the study of the radial (and vertical) metallicity gradients in the Galaxy. 

Planetary nebulae are the ejecta of evolved 1-8 \msun~stars, thus they represent a continuum of stellar populations from very old ages through relatively young, with ages $\sim$0 to $\sim$10 Gyr, which makes them very versatile probes for Galactic evolution. Gas-phase abundances of {$\alpha$-elements} observed through emission-lines and measured in their shell should be the same as their progenitors' at the time of formation, since these elements do not change considerably during the evolution of these stars. Thus, tracing the abundances of such elements is equivalent to probing the Galaxy's abundance with look-back time from $\sim$0 to $\sim$10 Gyr, depending on the initial progenitor mass of the PNe.

\subsection{Elemental abundances}
We use PN elemental abundances both for progenitor dating purposes, and to determine the metallicity gradients. We select the elemental abundances from all references included in \citet{SH18}, to whose we add the abundances from the data sets by \citet{2003MNRAS.345..186T}, \citet{Miller2019}, \citet{McNabb2016}, and \citet{DI2015}, which were not included therein. In the references where abundances are given both for the whole nebula and for different parts of the PN, we use only those measures based on spectra that include the whole PN, so abundances from different references are readily comparable with one another. The final abundances have been curated, i.e.,  we have recalculated the ionization correction factor (ICF) uniformly, as described in \citet{SH18}. 

In Table~\ref{table:abun}
we give the average abundances used in this paper. Column (1) gives the PN~G name, then in columns (2) through (4) we give the C, N, and O abundances, in the usual scale log(X/H)+12, averaged from all the available references. The uncertainties, given in dex, are the dispersion of that elemental abundance across the literature, estimated with $\sigma({\rm log(O/H)})=0.434\times[\sigma{\rm (O/H)/<O/H>}]$.
We also produced a curated catalog of selected PN abundances, shown in Table~\ref{table:abunsample}. For this selection we prioritize abundances from space-based observations unless they are deemed uncertain in the original papers; otherwise, we use the published ground-based abundances, prioritizing the most recent, or recently revised, ones. Curated abundances are given with their uncertainty, when available in the original reference. 

\begin{table*}
\caption{Averaged Elemental Abundances of Galacticv PNe}          
\label{table:abun} 
\begin{tabular}{lrrrl}
\hline
PN~G &  log(C/H)+12&  log(N/H)+12&  log(O/H)+12 & Prog. Age\\
\hline
002.7-04.8 &     9.02$\pm$0.20 &     8.68$\pm$0.27 &     8.68$\pm$0.20 & OPPN\\
006.1+08.3 &     8.58$\pm$0.20 &     7.92$\pm$0.17 &     8.57$\pm$0.20 & OPPN\\
010.1+00.7 &     8.23$\pm$0.20 &     8.65$\pm$0.20 &     8.26$\pm$0.20 & YPPN\\
\hline
\end{tabular}
\tablefoot{
Elemental abundances have been averaged from all references included in \citet{SH18}, \citet{2003MNRAS.345..186T}, \citet{Miller2019}, \citet{McNabb2016}, and \citet{DI2015}; the age of the PPN progenitor (OPPN=old, YPPN=young) is defined according to \citet{SH18} and \citet{2016MNRAS.460.3940V}, see text.}
\tablefoottext{The full table is available electronically at the CDS.}

\end{table*}

\begin{table*}
\caption{Selected Elemental Abundances of Galactic PNe}          
\label{table:abunsample} 
\begin{tabular}{lrrrl}
\hline
PN~G &  log(C/H)+12&  log(N/H)+12&  log(O/H)+12 & Prog. Age\\
\hline
002.7-04.8 &     9.02$\pm$0.20 &     8.68$\pm$0.20 &     8.48$\pm$0.20& OPPN\\
006.1+08.3 &     8.58$\pm$0.20 &     8.03$\pm$0.20 &     8.58$\pm$0.20 &OPPN\\
010.1+00.7 &     8.23$\pm$0.20 &     8.65$\pm$0.20 &     8.26$\pm$0.20 &YPPN\\
\hline
\end{tabular}
\tablefoot{In the computation of abundances, space-based observations and most recent ground-based abundances have been prioritized (see main text)}

\tablefoottext{The full table is available electronically at the CDS.}

\end{table*}
\subsection{Velocities of PNe Central Stars}

For a sizable sample of CSs whose PNe have a statistical distance in our catalog, Gaia DR3 provides equatorial proper motions $\mu_{\alpha*} \pm \sigma_{\mu_{\alpha*}}$\footnote{the asterisk, indicating that the longitudinal proper motion is measured along the great circle at the star's location, will be dropped and implicitly assumed in the following.}, $\mu_{\delta} \pm \sigma_{\mu_{\delta}}$; we use this kinematic information as insight for their populations and ages.
After converting DR3 proper motions and their standard deviations in galactic coordinates $\mu_{\rm l} \pm \sigma_{\mu_{\rm l}}$, $\mu_{\rm b} \pm \sigma_{\mu_{\rm b}}$ via the transformation matrices $A'_{\rm G}$ and $C_{\rm Gal}$ defined by Eq. 4.62 and 4.79 of the Gaia EDR3
\href{https://gea.esac.esa.int/archive/documentation/GEDR3/Data_processing/chap_cu3ast/sec_cu3ast_intro/ssec_cu3ast_intro_tansforms.html#SSS1}{online documentation},
we compute the corresponding observed spatial velocities plus errors, given in ${\rm [km~s^{-1}]}$, as
\begin{equation}
\begin{array}{l}
 V_{\rm l}=4.740470446\mu_{\rm l}\times D  \\
 \sigma_{V_{\rm l}}=4.7407470446\times D \sqrt{\mu_{l}^2\sigma_{\rm D}^2+\sigma_{\mu_l}^2} 
\end{array}
\end{equation}
 \begin{equation}
 \begin{array}{l}
    V_{\rm b}=4.74047\mu_{\rm b}\times D\\
    \sigma_{V_{\rm b}}=4.74074\times D \sqrt{\mu_{b}^2\sigma_{\rm D}^2+\sigma_{\mu_b}^2}
    \end{array}
   \end{equation}
where $l$ and $b$ are the Galactic longitude and latitude. We also adopt
$\sigma_{\rm D}\equiv (\sigma_{rm D^-}+\sigma_{\rm D^+})/2$. 
The third component of the velocity is the radial component, $V_{\rm r}$, which has been observed independently for $498$ PNe of our sample \citep{Durand1998}\footnote{Gaia DR3 had measured radial velocities for only 13 CSs of our Calibrator Sample}.

Having secured distances and kinematic data for these stars, we can estimate their {\it peculiar} velocity $V_{\rm pec}$ as the residual stellar motion which does not conform to the general Galactic rotation. To this end, we assumed an empirical model for the Galactic rotation curve which follows \citet{Eilers19} for galactocentric distances larger than 5 kpc, while for inner distances we made use of the results of \citet{B2014} 
-- who resorted to H~I and H~II regions, as well as CO emission lines, as tracers of the inner Galactic Disk -- adapting their Rotation Curve to the fundamental parameters found by \citet{Eilers19}, i.e., $R_{\odot}=8.122$ kpc, the galactocentric distance of the Sun, and $V_{\rm LSR}=229$ ${\rm [km~s^{-1}]}$,  the circular velocity of the Local Standard of Rest (LSR). We have combined these data into a grid of Galactic radii and corresponding rotation velocities, plus uncertainties, given in Table~\ref{table:Tabrot_merged}.
Moreover, in the following calculations, the velocity of the Sun with respect to the LSR has been set to ($U_{\odot}$, $V_{\odot}$, $W_{\odot}$)=(11.1, 12.4, 7.25) ${\rm [km~s^{-1}]}$, according to \citet{Scho10}. 
\begin{table}
\caption{Adopted Galactic Rotation Curve\tablefootmark{1}}          
\label{table:Tabrot_merged}    
\begin{tabular}{lrrl}
\hline
$R_{\rm G}$  &      $V_{\rm c}$  &  $R_{\rm G}$  &  $V_{\rm c}$  \\

 [kpc]&               [km/s]&         [kpc]&         [km/s]\\

\hline
1.62 &  $208.42^{+5.80}_{-5.80}$   &  13.74 &    $217.47^{+0.64}_{-0.51}$ \\ 	   
2.61 &  $206.23^{+1.60}_{-1.60}$    &  14.24 &   $217.31^{+0.77}_{-0.66}$ \\ 	   
3.55 &  $218.98^{+2.54}_{-2.54}$    & 14.74  &   $217.60^{+0.65}_{-0.68}$ \\ 	   
4.52 &  $238.93^{+2.20}_{-2.20}$   & 15.22  &    $217.07^{+1.06}_{-0.80}$ \\	   
5.27 &  $226.83^{+1.91}_{-1.90}$   & 15.74  &     $217.38^{+0.84}_{-1.07}$ \\ 	        
5.74 &  $230.80^{+1.43}_{-1.34}$   &16.24   &     $216.14^{+1.20}_{-1.48}$ \\ 	   
6.23 &  $231.20^{+1.70}_{-1.10}$   &16.74   &     $212.52^{+1.39}_{-1.43}$ \\ 	   
6.73 &  $229.88^{+1.44}_{-1.32}$   &17.25   &     $216.41^{+1.44}_{-1.85}$ \\	   
7.22 &  $229.61^{+1.37}_{-1.11}$   &17.75   &     $213.70^{+2.22}_{-1.65}$ \\ 	   
7.82 &  $229.91^{+0.92}_{-0.88}$   &18.24   &     $207.89^{+1.76}_{-1.88} $ \\	   
8.19 &  $228.86^{+0.80}_{-0.67}$   &18.74   &     $209.60^{+2.31}_{-2.77} $ 	\\   
8.78 &  $226.50^{+1.07}_{-0.95}$   &19.22   &     $206.45^{+2.54}_{-2.36}  $	 \\  
9.27 &  $226.20^{+0.72}_{-0.62}$   &19.71   &     $201.91^{+2.99}_{-2.26}  $	 \\  
9.76 &  $225.94^{+0.42}_{-0.52}$   &20.27   &     $199.84^{+3.15}_{-2.89}  $	\\   
10.26&  $225.68^{+0.44}_ {-0.40}$   &20.78   &   $198.14^{+3.33}_{-3.37}  $  \\ 
10.75&  $224.73^{+0.38}_ {-0.41}$   &21.24   &    $195.30^{+5.99}_{-6.50}  $ \\ 
11.25&  $224.02^{+0.33}_ {-0.54}$   &21.80   &    $213.67^{+15.38}_{-12.18}  $ \\
11.75&  $223.86^{+0.40}_ {-0.39}$   &22.14   &    $176.97^{+28.58}_{-18.57}  $ \\
12.25& $222.23^{+0.51}_ {-0.37}$   &22.73   &      $193.11^{+27.64}_{-19.05}  $ \\
12.74& $220.77^{+0.54}_ {-0.46}$   &23.66   &      $176.63^{+18.67}_{-16.74} $\\
13.23& $220.92^{+0.57}_ {-0.40}$   &24.82   &      $198.42^{+6.50}_{ -6.12}   $\\
\hline
\end{tabular}
\tablefoot{
\tablefoottext{1}{Derived from \citet{Eilers19} and \citet{B2014}}
}
\end{table}

We first transform the observed spatial velocities into cartesian components along the {\it local triad} at the star as
\begin{equation}
{\bf v}_{rel}= \begin{pmatrix} {\bf e}_l & {\bf e}_b & {\bf e}_r \end{pmatrix} \cdot \begin{pmatrix} V_l \\ V_b \\ V_r \end{pmatrix} \equiv A \cdot \begin{pmatrix} V_l \\ V_b \\ V_r \end{pmatrix}
\end{equation}
where \\
${\bf e}_l^T=(-\sin l,\cos l,0)$, ${\bf e}^T_b=(-\sin b \cos l, -\sin b \sin l, \cos b)$, ${\bf e}_r^T=(\cos b \cos l, \cos b \sin l, \sin b)$, and {\it A} is the so-called direction cosine matrix representing the rotation to the local frame.

To derive galactocentric velocities we must account for the Sun's galactocentric velocity, i.e. ${\bf v}_\odot=(U_\odot, V_\odot+V_{LSR}, W_\odot)$, so that
\begin{equation}
{\bf v}_{gal}=\begin{pmatrix} U \\ V \\  W\end{pmatrix}={\bf v}_{rel}+{\bf v}_\odot
\end{equation}
For the following analysis, it is useful to express spatial velocities in cylindrical coordinates, for which we need to compute the galactocentric azimuth $\phi$ as \citep[e.g.][]{2019A&A...621A..48L}
\begin{equation}
\begin{array}{l}
    X=R_{\rm \odot}-D\cos{l}\cos{b}\\
    Y=D\sin{l}\cos{b}\\
    \phi=\tan^{-1}{Y/X}
    \end{array}
\end{equation}
where $D$ is the star's heliocentric distance and $\phi$ is taken as positive in the direction of galactic rotation; the radial, azimuthal, and vertical components of the spatial velocity are then given by:
\begin{equation}
\begin{array}{l}
V_{\rm R} =-U\cos\phi+V\sin\phi \\
V_\phi=U\sin\phi+V\cos\phi \\
V_{\rm z} = W
\end{array}
\end{equation}

Finally, we are able to estimate the peculiar velocity of each star by subtracting from the azimuthal component $V_\phi$ the circular motion due to differential galactic rotation. To do so, for each PN we 
 compute the galactocentric radius $R_{\rm G}$ projected onto the Galactic plane as:
\begin{equation}
    R_{\rm G}=\sqrt{R_\odot^2+D^2\cos^2{b}-2R_\odot D\cos{l}\cos{b}},
\end{equation}
and 
interpolate linearly between the nearest two values of $R_{\rm G}$ of Table~\ref{table:Tabrot_merged} to calculate the star's circular velocity $V_{\rm c}(R_{\rm G})$.

The absolute space motion of the stars (or the PNe), which we call $V_{\rm pec}$, can be obtained by quadrature of the three velocity components as 
\begin{equation}
\begin{array}{l}
V_{\rm pec}=\sqrt{V_{\rm R}^2+V_{\rm z}^2+(V_\phi-V_{\rm c})^2} \\
\sigma_{V_{\rm pec}} \approx \sqrt{\sigma_{V_{\rm l}}^2+\sigma_{V_{\rm b}}^2+\sigma_{V_{\rm r}}^2}    
\end{array}
\label{eq:peculiar}
\end{equation}
where $\sigma_{V_{\rm pec}}$ is a lower-limit uncertainty, which assumes uncorrelated variables and perfect model. Peculiar velocities for the PNe are reported in Table~\ref{table:catalog}.

\begin{figure}
   \centering
   \includegraphics[width=8cm]{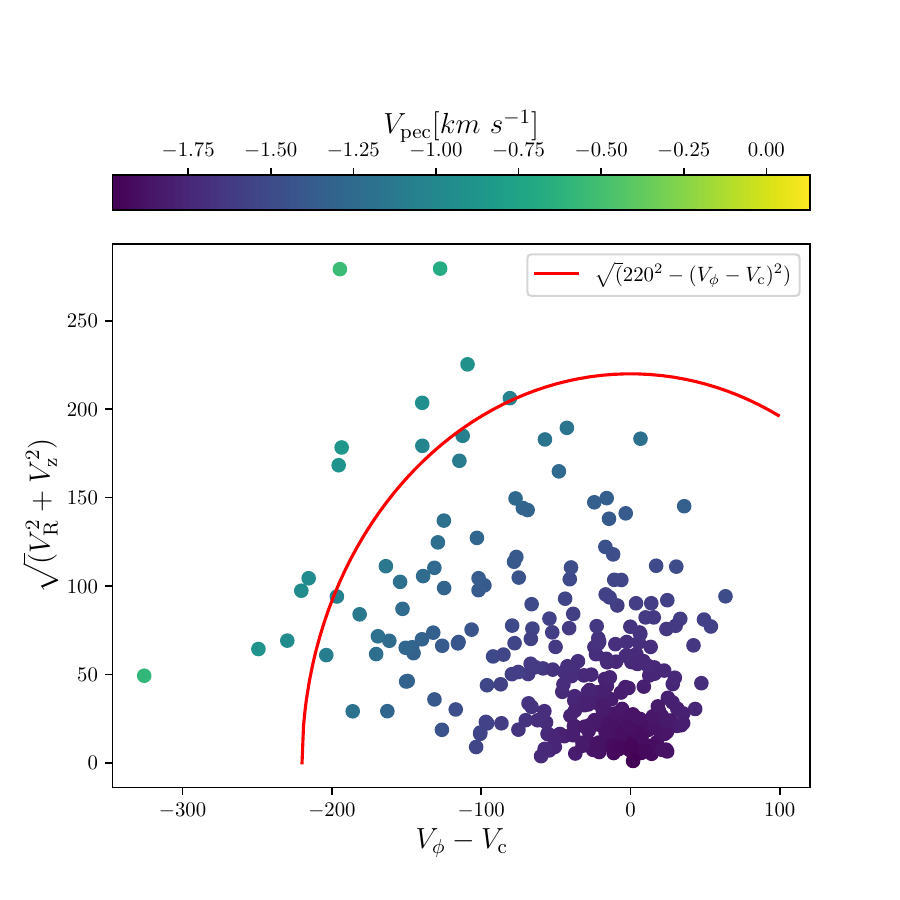}
  
   \caption{Halo PNe selected through the velocity analysis of the CSs, by means of the Toomre diagram. The colour bar is coded for the peculiar velocities as in Eq. ~\ref{eq:peculiar}. The dots outside the region encircled by the red curve correspond to Halo PNe.}
              \label{fig:toomre}
    \end{figure}

In Figure~\ref{fig:toomre} we put the inferred peculiar velocities in the so-called Toomre diagram, which can be used to identify regions of high probability of finding Disk versus Halo stars. In particular, following \citet{2017ApJ...845..101B},
we use  $V_{\rm pec}> 220$ ${\rm [km~s^{-1}]}$ as threshold value for rejecting Halo stars.

    \begin{figure}
   \centering
   \includegraphics[width=8cm]{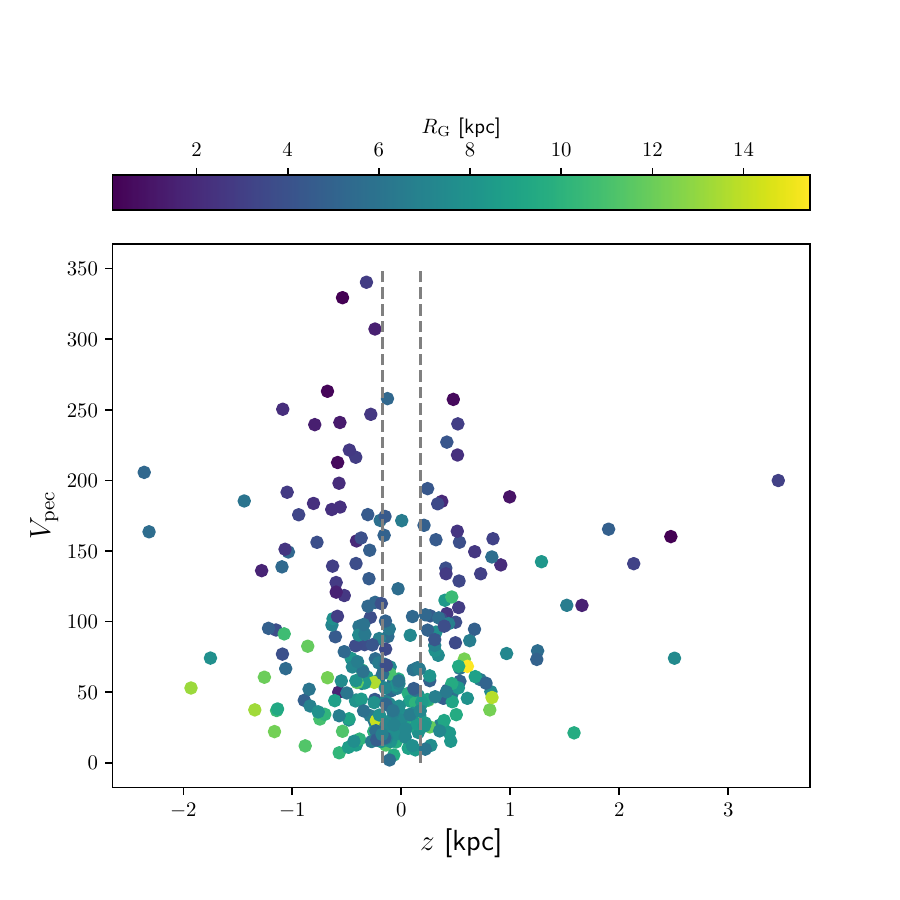}
  
   \caption{Galactic distribution of $V_{\rm pec}$ from Eq. ~\ref{eq:peculiar}, with Galactic altitude, where the colour is coded for the distance from the Galactic center $R_{\rm G}$. The vertical thin lines encompass $|z|<0.175$ zone .}
              \label{fig:galaxy}
    \end{figure}

\subsection{Galactic PN Populations, and Dating PN progenitors}

We derive heliocentric distances $D$ and their confidence limits, as described in subsection 2.3, for 843 Galactic PNe. The range of distances spanned by these PNe is $\sim 150$ to $\sim$27,000 [pc], and most of them can be included in a  general Disk population, thus they are adequate for our science scope. We populate the Halo PN sample selecting them by peculiar velocity, where Halo PNe have $V_{\rm pec}>220$ [km s$^{-1}]$. This selection yields to 13 Halo PNe. 
We also tentatively define the Bulge population similarly to \citet{SH18}, by selecting those PNe in the central 3$\times$3 [kpc], thus with  $|z|<3$ and $0<R_{\rm G}<3$ [kpc]. To constrain the Bulge sample even more we select in this sample only those PNe whose apparent radius is $\theta<5$ [arcsec]. Our selection gives 102 Bulge PNe. Whether a PN belongs to the Halo or Bulge populations, according to our selection, is noted in Table~\ref{table:catalog}.

In Figure~\ref{fig:galaxy} we plot the distribution of the peculiar velocity vs. $z$, where the colour indicates the distance from the Galactic center. We observe that in the area between the two grey lines ($\pm$0.175 kpc) PNe with high peculiar velocity are away from the Galactic center. It is worth noting that all Bulge PNe selected as in $\S$3.3 are outside the grey lines, and have intermediate $V_{\rm pec}$.

   \begin{figure}
   \centering
   \includegraphics[width=8cm]{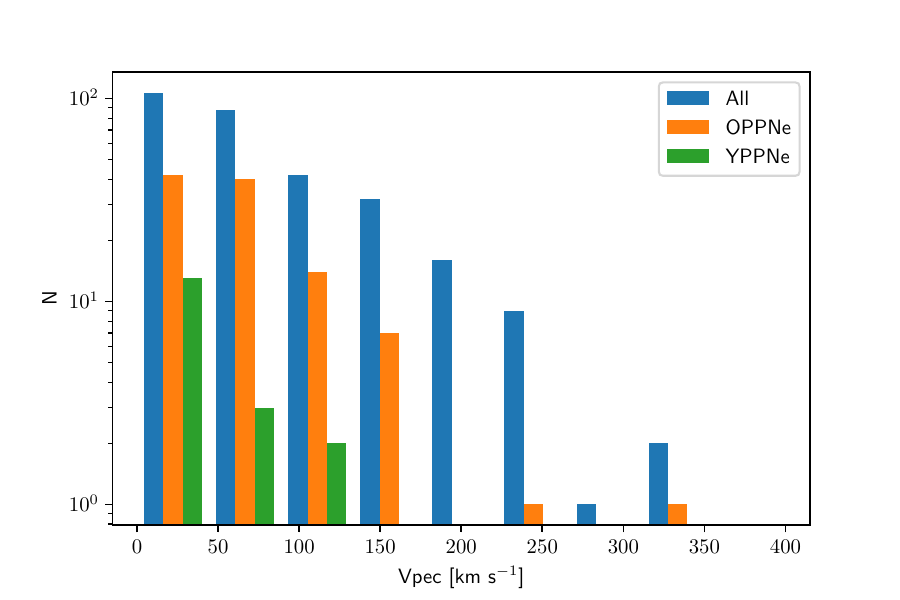}
      \caption{Histogram of the spatial peculiar velocity $V_{\rm pec}$ ${\rm [km~s^{-1}]}$ for the general Disk population, and for the OPPNe and YPPNe separately. The velocity bin is 10 km s$^{-1}$. The y-axis is in logarithmic scale.}
         \label{fig:histo}
 
   \end{figure}

Dating PN progenitors is not trivial, and it has been attempted by Peimbert and Torres-Peimbert, by dividing Galactic PNe into populations (i.e., PN Types I through V). This approach utilizes both the chemistry of the PNe and their kinematics. \citet{SH18} used the final yields of the {\it evolutionary} elements, compared to nebular abundances of carbon and nitrogen to select two PN's progenitor populations with extreme ages, called OPPNe (old progenitors PNe, with progenitor ages$>$7.5 Gyr), and YPPNe (young progenitors PNe, with progenitor ages $<$ 1 Gyr). This selection is based on the observation that PNe with old and young progenitors occupy markedly different loci on the N/H versus C/H, and N/H vs. O/H, planes. Fig. 2 of \citet{2016MNRAS.460.3940V} was used to determine that OPPNe have C/H $>$ N/H or [log(N/H) + 12] $<$ 0.8 × [log(O/H) + 12] + 1.4, and YPPNe have C/H $<$ N/H or [log(N/H) + 12] $>$ 0.6 × [log(O/H) + 12] + 3.3. In  Tables~\ref{table:abun}  and \ref{table:abunsample}, column 5, we give the PN progenitor age group as derived above, when available. 

More insight on dating PN progenitors are derived from the comparison  of Gaia proper motions described in the previous sub-section, with the ages derived purely from the PN chemistry above.
In Figure~\ref{fig:histo} we show the resulting distribution of $V_{\rm pec}$ for the general PN population (which includes OPPNe, YPPNe, as well as the intermediate age progenitor PNe, and those PNe whose age from chemistry was unavailable), then by plotting separately OPPNe and YPPNe. We find that the OPPNe distribution peaks at $V_{\rm pec}\sim 60$ ${\rm [km~s^{-1}]}$),  while the YPPNe distribution peaks at a smaller velocity ($\sim 25-30$ ${\rm [km~s^{-1}]}$). This plot shows the correlation between completely independent dating systems, from the chemistry and the kinematics of the PNe, giving us confidence about using the OPPNe and YPPNe distributions.

\subsection{Galactic metallicity gradients and other metallicity variations}
The radial metallicity gradients in this paper have been calculated by orthogonal distance regression, which makes use of the uncertainties in the measured quantities $R_{\rm G}$ and ${\rm log(O/H)}+12$.
The uncertainties in the elemental abundances are described in $\S$3.1, where we indicate which literature fonts we chose the abundances from, and how errors are estimated both in the average and chosen abundances (see Table~\ref{table:abun}). The uncertainties in $R_{\rm G}$ are computed by first-order propagation of the symmetrized error of the distance scale, $\sigma_{\rm D}=(\sigma_{\rm D^-}+\sigma_{\rm D^+})/2$, thus obtaining $\sigma_{R_{\rm G}}=|D\cos^2{b}-R_\odot\cos{l}\cos{b}|\cdot\sigma_D/R_{\rm G}$. The regression yields to a slope and an intercept, along with their formal errors.

\begin{figure}
   \centering
   \includegraphics[width=8cm]{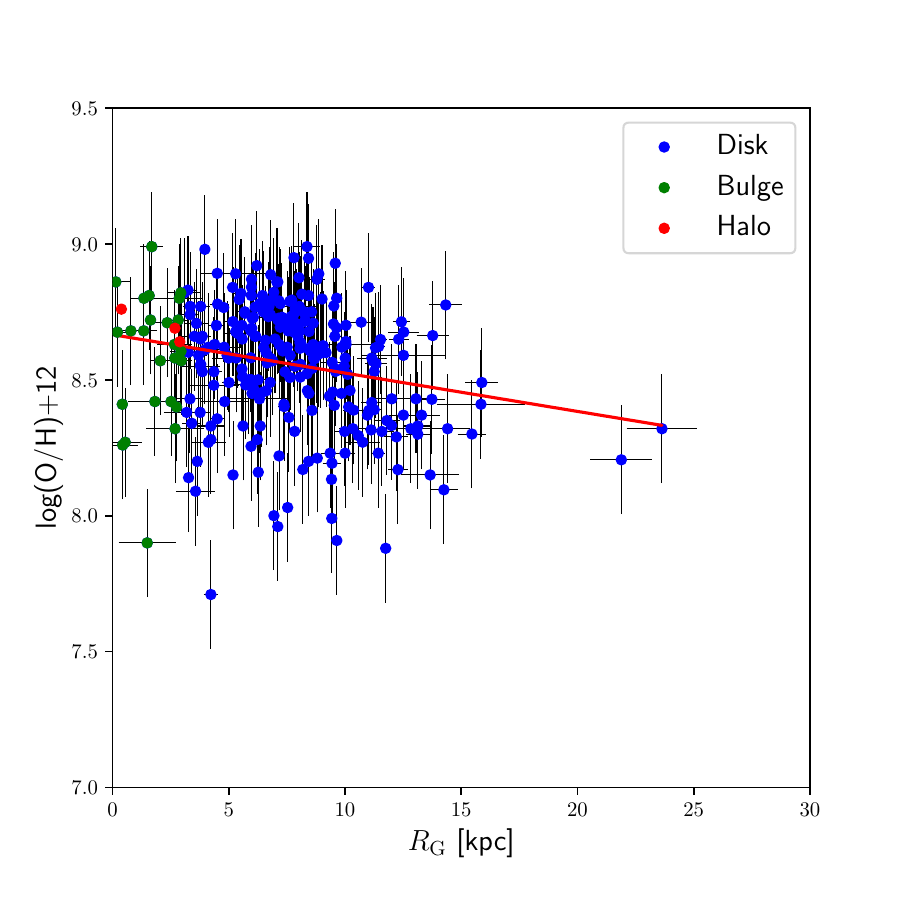}
   \caption{Radial oxygen gradients for the complete PN sample for which we have a statistical distance and at least one O/H measurement in the literature. Oxygen abundances are averages of all published abundances, and the error bars represent the dispersion. All PNe have been included in these plots, and Halo and Bulge populations are indicated with different colours. The red line is the linear fit to the data; slopes and intercepts for the different populations are in Table~\ref{table:gradients}.}
              \label{fig:grad_all}
    \end{figure}
    
\begin{figure}
   \centering
   \includegraphics[width=8cm]{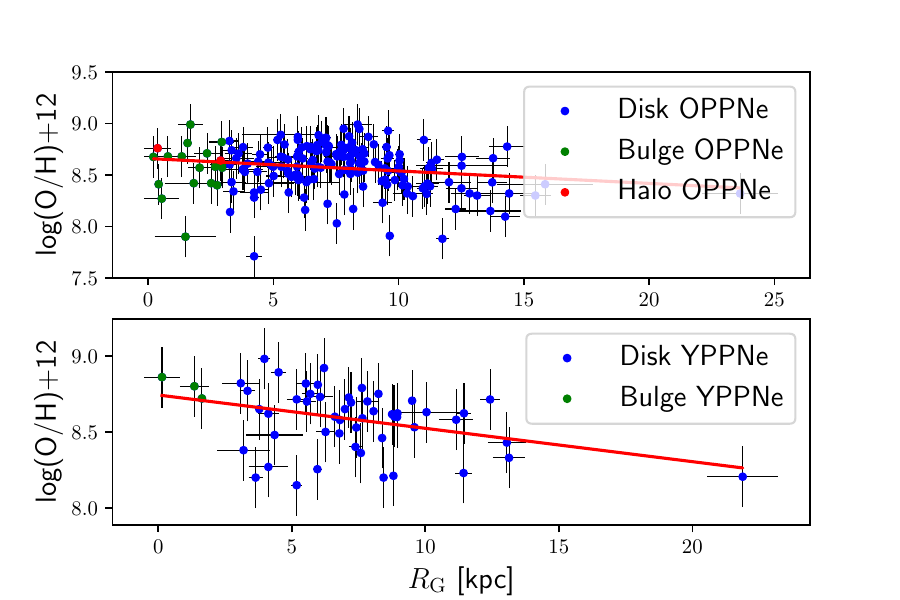}
   \caption{Radial oxygen gradients for the OPPN (top) and the YPPN (bottom) populations. All PNe have been included in these plots, and Halo and Bulge populations are indicated with different colours. Oxygen abundances are the selected ones here. The red lines are the linear fits of the data; slopes and intercepts for the different populations are in Table~\ref{table:gradients}.}
              \label{fig:grad_age}
    \end{figure}

In Table~\ref{table:gradients} we present the estimation of the gradient slopes obtained using different subsets of PNe, and with both averaged and selected chemical abundances. We utilize all PNe where at least one oxygen measurement was available in the literature, excluding only the extreme outliers (log(O/H)+12$<7.5$ and log(O/H)+12$>9$).
All radial oxygen gradients that we derive in this paper have negative slopes. The general PN population (Fig.~\ref{fig:grad_all} has gradient slope of -0.0144$\pm$0.00385 [dex kpc$^{-1}$] when we use the average abundances. By excluding the Halo population (2 PNe??) the gradient does not change significantly. 
\begin{table*}
\caption{Radial Oxygen Gradients}          
\label{table:gradients}     
\begin{tabular}{lrrr}
\hline
Sample&  N& slope& intercept\\
& &  [dex kpc$^{-1}$]& log(O/H)+12\\
\hline
Average abundances\\
\hline
All&  288& -0.0144$\pm$0.00385& 8.669$\pm$0.0307\\
Excluding Halo PNe& 285& -0.0141$\pm$0.00390& 8.666$\pm$0.0314\\
OPPNe&  186& -0.0121$\pm$0.00465& 8.660$\pm$0.0376\\
YPPNe& 55& -0.0220$\pm$0.00758& 8.743$\pm$0.0601\\

\hline
Selected abundances\\
\hline
All&  288&  -0.0163$\pm$0.00385& 8.672$\pm$0.0312\\
Excluding Halo PNe& 285& -0.0159$\pm$0.00392& 8.668$\pm$0.0319\\
OPPNe&  186& -0.0132$\pm$0.00464& 8.659$\pm$0.0373\\
YPPNe& 55& -0.0278$\pm$0.00789& 8.759$\pm$0.0629\\
\hline
\end{tabular}
\tablefoot{Gradients are estimated using chemical abundances from Tables~\ref{table:abun} (Average abundances) and \ref{table:abunsample} (Selected abundances), and for the listed PN samples, of numerosity N.}

\end{table*}

For all comparable samples, the evolution goes in the direction of gradients steepening with time since Galaxy formation. In fact, for the OPPN and YPPN populations we 
derive gradient slopes $\Delta{\rm log(O/H)}/\Delta R_{\rm G}=-0.0121\pm0.00465$ 
and -0.022$\pm$0.00758 [dex kpc$^{-1}$] respectively when we use average abundances, and  $\Delta{\rm log(O/H)}/\Delta R_{\rm G}=-0.0132\pm0.00464$ 
and -0.0278$\pm$0.00789 [dex kpc$^{-1}$] respectively if we use the selected abundances (plots in Fig.~\ref{fig:grad_age}). This result indicates a mild -- but significant when compared to the uncertainties -- evolution of the metallicity gradient with time. Our results point to a steeper gradient slope for the younger stellar population than for the older stellar population, with slope difference of $\sim$0.15 dex. This implies a slow time evolution of the gradient. The time evolution of the gradient slope does not vary whether or not we include Halo PNe in the OPPN sample (see Table~\ref{table:gradients}).

It is beyond the scope of this paper to discuss the causes for gradient steepening. Chemical evolutionary models of star forming galaxies \citep{2013A&A...554A..47G} indicate that steepening with time of the radial metallicity gradients occurs when the chemical evolution proceeds with feedback of fresh, metal poor gas from the environment. Such models can not predict the gradient ab initio, thus, metallicity gradients for old populations such as those from OPPNe represent a precious constraint.


Finally, we tested the oxygen evolution (or chemical enrichment) by comparing average oxygen abundances in the OPPNe and YPPNe samples, to derive an average Galactic enrichment of $\sim$0.06 dex between the age$>$7.5 Gyr and the age$<$1 Gyr populations.

\section{Results and conclusions}

This paper yields essentially to three novel results for PNe in the Galactic environment. First, we produced a new distance scale based on DR3 parallaxes, which is given in Eq.\ref{adopted}.  Distances derived from the new scale are published as a catalog in this paper (Table~\ref{table:catalog}). By comparing our new distances with the DR3 CS parallaxes, as in the analysis by \citet{Smith2015}, we find $<K>=0.964$, where $K=D\times \varpi_{\rm c}$, and $<\sigma_{\rm K}>=0.154$ if we exclude from the comparison the PNe with ${\rm log}M_{\rm i}<-2$, and  $<K>=0.931$ if we include all calibrators. Comparing these results with those of Paper I, where the best scale obtained with DR2 parallaxes gave $<K>=0.948$ and $<\sigma_{\rm K}>=0.25$, we conclude that the use of DR3 parallaxes as calibrators successfully improved upon the Galactic PN distance scale. 

We have calculated $<K>$ based on DR3 parallaxes for the most commonly used PN distance scales, such as \citet{F16} (hereafter FPB) and \citet{SSV} (hereafter SSV). We found $<K_{\rm FPB}>= <D_{\rm FPB}\times \varpi_{\rm c}> = 1.272$ and $<K_{\rm SSV}>= <D_{\rm SSV}\times \varpi_{\rm c}>= 1.203$ with the exclusion of calibrators with ${\rm log}M_{\rm i}<-2$, and $<K_{\rm FPB}>=1.36$ and $<K_{\rm SSV}>=1.29$  when including all calibrators. We conclude that the scale presented in Eq.\ref{adopted} gives the most reliable distances to date, and distances in Table~\ref{table:catalog} should be used, unless an accurate parallax measurement of the CS or PN under study is available, from DR3 or otherwise.
Moreover, our analysis revealed that the ${\rm log}R-{\rm log}S_{\rm H\beta}$ distance scale of Galactic PNe  has an intrinsic mean dispersion of $\approx 0.1$ dex around the $\log R-\log S_{\rm H\beta}$ linear relation, which can be accounted for by the variation of the PN ionized mass (excluding objects with ${\rm log}M_{\rm i} < -2$), bringing us to the conclusion that statistical distance scales such as this can hardly be improved by increasing the accuracy of calibrators.

Second, we used Gaia DR3 proper motion to have a thorough description of the peculiar velocities of the CSs, or the PNe, and to characterize the Halo population. The peculiar velocity derivation has the additional advantage to be a completely independent assessment of the progenitor ages of the PNe, thus validating even further the OPPN and YPPN classes based on elemental abundances rations C/O and N/O. The proper motion and peculiar velocity analysis of $\S$~3.2 provided a tool to exclude Halo PNe from the gradient analysis. Even more importantly, it gave us additional confidence in the progenitor dating from chemical analysis, as described in $\S$~3.3. This is, to our knowledge, the first time that dating PNe is consistent with two completely independent approaches, while in the past the dating scheme has used either approach, or both approaches for different age classes, such as in defining Peimbert's PN Types \citep{1978IAUS...76..215P}. Dating PN progenitors is essential to study the time evolution of the radial metallicity gradients in the Milky Way.

Third, we measure radial oxygen gradients based on DR3 distance scale, with Halo PN selected through their kinematic properties measured by Gaia DR3, and progenitor ages from chemical evolutionary assessment, confirmed via the peculiar velocities determined by Gaia. The negative, shallow gradients that we determine, and the mild evolution of the gradients -- steepening since the formation of the Galaxy--- are a confirmation of past results both in the Galaxy \citep{SH18} and in the nearby spiral galaxies examined \citep{2019RMxAA..55..255P,2016A&A...588A..91M}, where all emission-line probes in star-forming galaxies (including the Milky Way) indicate that the oxygen radial gradients are steeper in the younger populations. For the gradients we have used curated catalogs of published abundances, which we give in Tables~\ref{table:abun} and \ref{table:abunsample}. 

It is worth noting that, while our results confirm most of the metallicity gradients measurements based on gas-phase data published to date, recent cluster and stellar results show a more controversial picture \citep{2017A&A...600A..70A,2023A&A...669A.119M,2023A&A...674A..38G}. Stellar metallicity are mainly inferred from iron rather than oxygen; since iron atoms in PNe are mostly found in condensed state \citep[e.g.][]{DI2015}, a direct comparison between the stellar and nebular samples is typically problematic. 

The work of \cite{2023A&A...669A.119M} also include $\alpha$-element gradients, indicating minimal oxygen gradient evolution, compatible with flattening of gradients with time. Their oxygen gradient slopes of the age$<$1 Gyr and age$>$3 Gyr bins in their table A.10. are the same within the uncertainties. We underline that these results are not incompatible with ours, since we are comparing different age ranges: our PN progenitors are in the age$<$1 Gyr and age$>$ 7.5 Gyr bins, whereas their oldest clusters are within 7 Gyr. Furthermore, migration, and evolution due to inflow, of clusters and stars may be different in ways that are unpredictable when measuring radial gradients.

Interestingly, \citep{2022MNRAS.517.2343B} who studied the radial oxygen gradients in M31, 
found that young, thin disk PNe describe a steeper radial gradient than the older, thick disk population, in qualitative agreement with our results, where radial metallicity gradients steepen with galaxy evolution. 
Quantitatively, in their Fig. 13 they attempt to unify the gradients slopes across galaxies, taking into account their scale lengths, and show the remarkable similarities between oxygen gradients in the Galaxy and M31.

More inspection of Galactic gradients from data and modeling comes from the recent analysis in \citet{2023NatAs.tmp..130L}, where they use integrated radial metallicities to find (see their Fig.~1) that radial Fe/H gradients from older populations are flatter than for the young population, in agreement with our results. They also compared the Galactic gradient from present-day stellar population with that obtained from gas-phase oxygen (H~II region), which is consistent with the PNe abundance tracers, reinforcing the confidence on our findings.

The distance scale that we presented here, and the distances in our catalog, will be used in the future to study the behaviour of dust and other elements in Galactic PNe in the framework of the chemical evolution of the Galaxy.

\begin{acknowledgements}
This work has made use of data from the European Space Agency (ESA) mission
{\it Gaia} (\url{https://www.cosmos.esa.int/gaia}), processed by the {\it Gaia}
Data Processing and Analysis Consortium (DPAC,
\url{https://www.cosmos.esa.int/web/gaia/dpac/consortium}). Funding for the DPAC
has been provided by national institutions, in particular the institutions
participating in the {\it Gaia} Multilateral Agreement.
B.B. acknowledges the support of the Italian Space Agency
(ASI) through contract 2018-24-HH.0 and its addendum 2018-24-HH.1-2022
to the National Institute for Astrophysics (INAF).\\

The authors would like to thank the anonymous referee whose comments and suggestions helped improving the original manuscript.
\end{acknowledgements}



\bibliographystyle{aa}
\bibliography{bibliopapPN.bib}

\end{document}